# Autoignition of Butanol Isomers at Low to Intermediate Temperature and Elevated Pressure

Bryan Weber[1], Kamal Kumar[2] and Chih-Jen Sung[3]
*University of Connecticut, Storrs, CT, 06269*

**Autoignition delay experiments for the isomers of butanol, including *n*-, *sec*-, *tert*-, and *iso*-butanol, have been performed using a heated rapid compression machine. For a compressed pressure of 15 bar, the compressed temperatures have been varied in the range of 725–855 K for all the stoichiometric fuel/oxidizer mixtures. Over the conditions investigated in this study, the ignition delay decreases monotonically as temperature increases and exhibits single-stage characteristics. Experimental ignition delays are also compared to simulations computed using three kinetic mechanisms available in the literature. Reasonable agreement is found for three isomers (*tert*-, *iso*-, and *n*-butanol).**

## Nomenclature

| | | |
|---|---|---|
| $P_C$ | = | compressed pressure |
| $P(t)$ | = | pressure as a function of time |
| $P'(t)$ | = | time derivative of pressure as a function of time |
| $\phi$ | = | equivalence ratio |
| $RCM$ | = | rapid compression machine |
| $\tau$ | = | ignition delay |
| $T_C$ | = | compressed temperature |
| TDC | = | top dead center |
| $X$ | = | mole fraction |

## I. Introduction

Recent instability in energy markets, as well as environmental concerns, have pushed a renewed interest in alternative sources of energy. For certain industries, and especially in the transportation sector, alternative fuels such as ethanol are replacing traditional petroleum-based fuels. Unfortunately, ethanol is generally considered to be a poor replacement for current fuels[1].

To help alleviate the concerns about the use of ethanol, a new generation of alternative fuels is being developed. One fuel of particular recent interest is butanol. *n*-Butanol has received significant attention as a transportation fuel with the potential to replace ethanol and even gasoline. In addition to *n*-butanol, there are three other isomers of butanol (*sec*-, *iso*-, and *tert*-butanol) that are being investigated as high octane gasoline additives[2]. The butanol system also comprises the smallest alcohol system with primary, secondary, and tertiary alcohol groups; therefore, studying the butanol system will provide a base from which to build models of higher alcohols.

Much of the recent work on the butanol system has focused on *n*-butanol. Fundamental combustion data such as laminar flame speeds, species sampling measurements, and ignition delays have been reported for *n*-butanol (cf. Refs. 3-7), but data are scarcer for the isomers. There has been only one study of the ignition delay of the isomers, conducted in a shock tube at high temperature and relatively low pressure[8]. Other workers have reported induction times[9], laminar flame speeds of the isomers[10-12], species profiles in non-premixed[13,14] and premixed[15] flames, and species profiles from a pyrolysis experiment[16].

Large gaps exist in the available data for the isomers of butanol. In particular, there is only one study of autoignition delays and that at relatively high temperature and low pressure conditions. To fully understand the combustion properties of the butanol system, it is imperative to have data over extensive variations in the range of thermodynamic parameters. With this in mind, this study presents autoignition delay data for all four isomers of butanol, at an elevated pressure of 15 bar and relatively low temperatures between 725 K and 855 K, collected using a rapid compression machine. Experimental results are also modeled using several kinetic mechanisms available in the literature.

[1] Research Assistant, Department of Mechanical Engineering University of Connecticut, Student Member AIAA.
[2] Research Assistant Professor, Department of Mechanical Engineering University of Connecticut.
[3] Professor, Department of Mechanical Engineering University of Connecticut, Associate Fellow AIAA.

## II. Experimental and Computational Specifications

### A. Rapid Compression Machine

Autoignition delay measurements are performed in a rapid compression machine (RCM). The RCM compresses a fixed mass of reactive mixture in approximately 25–35 milliseconds, using a pneumatically driven and hydraulically stopped piston. The piston is machined with crevices designed to suppress the roll up vortex and to provide a homogeneous reaction zone in the reaction chamber. The initial pressure, initial temperature, stroke of the piston, and clearance at top dead center (TDC) are varied to study different compressed temperature and pressure conditions. Further details of the RCM used in this study can be found in Ref. 17.

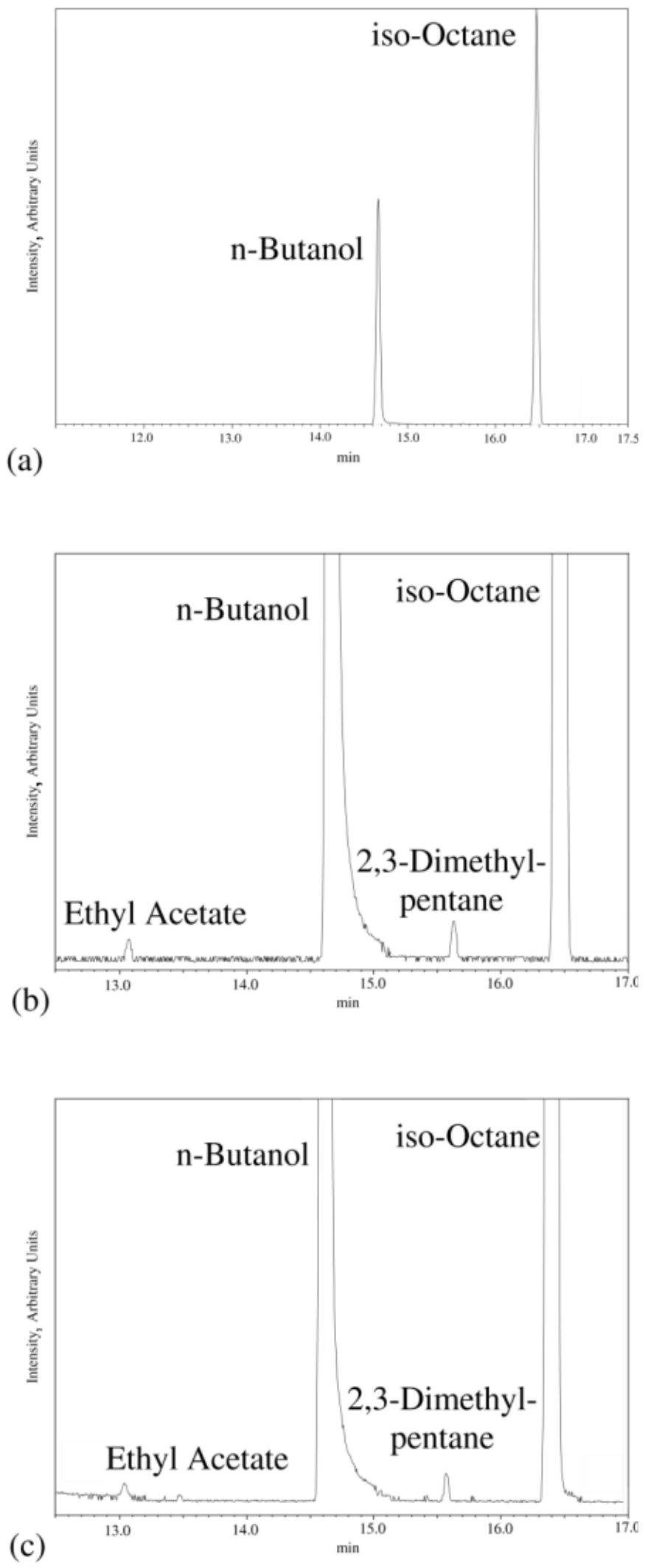

**Figure 1. Chromatograms of *n*-butanol and *iso*-octane separation.**
**(a) Overall chromatogram.**
**(b) Sample withdrawn from mixing tank.**
**(c) Liquid calibration sample.**

### B. Mixture Preparation

Mixtures are prepared in a stainless steel vessel equipped with a magnetically powered vane stirrer. The mixture composition is determined by specifying the mass of fuel, equivalence ratio ($\phi$), and oxidizer ratio ($X_{O2} : X_{N2}$, where $X$ indicates mole fraction). *n*-Butanol (anhydrous, 99.9%), *iso*-butanol (99.5%), *sec*-butanol (99.5%), and *tert*-butanol (99.7%) are used as fuels, while $O_2$ (99.8%) and $N_2$ (99.998%) are used to create the oxidizer. *n*-, *iso*-, and *sec*-butanol are liquids at room temperature and have relatively low vapor pressure, so they are massed gravimetrically in a syringe to within 0.01 g of the specified value. *tert*-Butanol is a solid at room temperature and is first melted in a glass container before being massed in the same manner as the rest of the fuels. Proportions of the gases in the mixture are determined manometrically and added at room temperature. The saturation vapor dependence of the fuels is taken from the *Chemical Properties Handbook* by Yaws[18]. The preheat temperature of the mixing tank is set above the saturation temperature of the fuels to ensure complete vaporization. The magnetic stirrer is activated, and the temperature of the mixture is allowed approximately 1.5 hours to reach steady state.

### C. Mixture Composition Check

Tests with Gas Chromatography/Mass Spectrometry (GCMS) are conducted to check that the expected mixture is present in the mixing tank for the entire duration of the experiments. A mixture is prepared as described previously, except a known concentration of *iso*-octane is added. This functions as an internal standard from which the concentration of fuel is calculated. As a demonstration, the following is a test performed with *n*-butanol.

Figure 1(a) shows the overall chromatogram of the separation of a sample withdrawn from the mixing tank, while Fig. 1(b) is an enlargement of Fig. 1(a) from 12.5 to 17 minutes. Figure 1(c) shows the chromatogram of a reference sample prepared from known masses of liquid *n*-butanol and *iso*-octane diluted in acetone. Comparing Figures 1(b) and 1(c) will reveal any thermal decomposition of the fuel during its preheat in the mixing tank. Since there are no peaks in Fig. 1(b) that are not also present in Fig. 1(c), it is concluded that there is no thermal decomposition of the fuel in the mixing tank. The two smaller peaks in Figs. 1(b) and 1(c) are impurities present in the *n*-butanol and *iso*-octane, ethyl acetate from *n*-butanol and 2,3-dimethylpentane from *iso*-octane. During this mixture composition check, the mixing tank is heated to 87 °C and contains a $\phi$=0.5 mixture in air, with 1.1% by mole *iso*-octane replacing an equivalent amount of nitrogen. In addition, the concentration of *n*-butanol in the mixing tank is calculated by using *iso*-octane as an internal standard. The response factor of *n*-butanol relative to *iso*-octane in the liquid sample is calculated based on the peak area ratio in Fig. 1(c) and the known concentrations of each component[19]. A total of five samples are withdrawn from the mixing tank and analyzed using the GCMS; the concentration of *n*-butanol is within 4% of the expected value for this representative case. Based on these results, it is concluded that the previously described mixture preparation technique is sufficient to obtain a homogeneous mixture.

### D. Experimental Conditions

Experiments are carried out at the same pressure and equivalence ratio condition for all four isomers of butanol. The compressed pressure ($P_C$) condition is chosen to provide data at engine relevant conditions, in a range that has not been covered previously. All experiments are carried out at $P_C$=15 bar, for $\phi$=1.0 mixture in nitrogen-oxygen air. The corresponding reactant mole fractions are: $X_{fuel} = 0.0338$, $X_{O2} = 0.2030$, and $X_{N2} = 0.7632$. The compressed temperature ($T_C$) conditions are similar for all the fuels, ranging from 725 K to 855 K. To the authors' knowledge, this is the first study of the autoignition of the butanol isomers in this pressure and temperature range.

### E. Experimental Reproducibility

Each compressed pressure and temperature condition is repeated at least six times to ensure reproducibility. The mean and standard deviation of the ignition delay for all concurrent runs is calculated; as an indication of reproducibility, one standard deviation of the ignition delays is less than 10% of the mean in all cases. Representative experimental pressure traces for simulations and plotting are chosen as the run whose ignition delay is closest to the mean. Furthermore, each new mixture preparation is checked against previously tested conditions to ensure consistency.

### F. Simulations and Determination of Compressed Temperature

Two types of simulations are performed using CHEMKIN-PRO[20]. The first type is a constant volume, adiabatic simulation, whose initial conditions are set to the pressure and temperature in the reaction chamber at top dead center (TDC). The second type includes the compression stroke and post-compression event by controlling the simulated reactor volume as a function of time. Heat loss during and after compression are modeled empirically to fit the experimental pressure trace of the corresponding non-reactive pressure trace, as described in Refs. 17 and 21-25. A non-reactive pressure trace is obtained by replacing oxygen with nitrogen in the mixture. This replacement maintains a similar mixture specific heat ratio, while eliminating oxidation reactions that can cause major heat release.

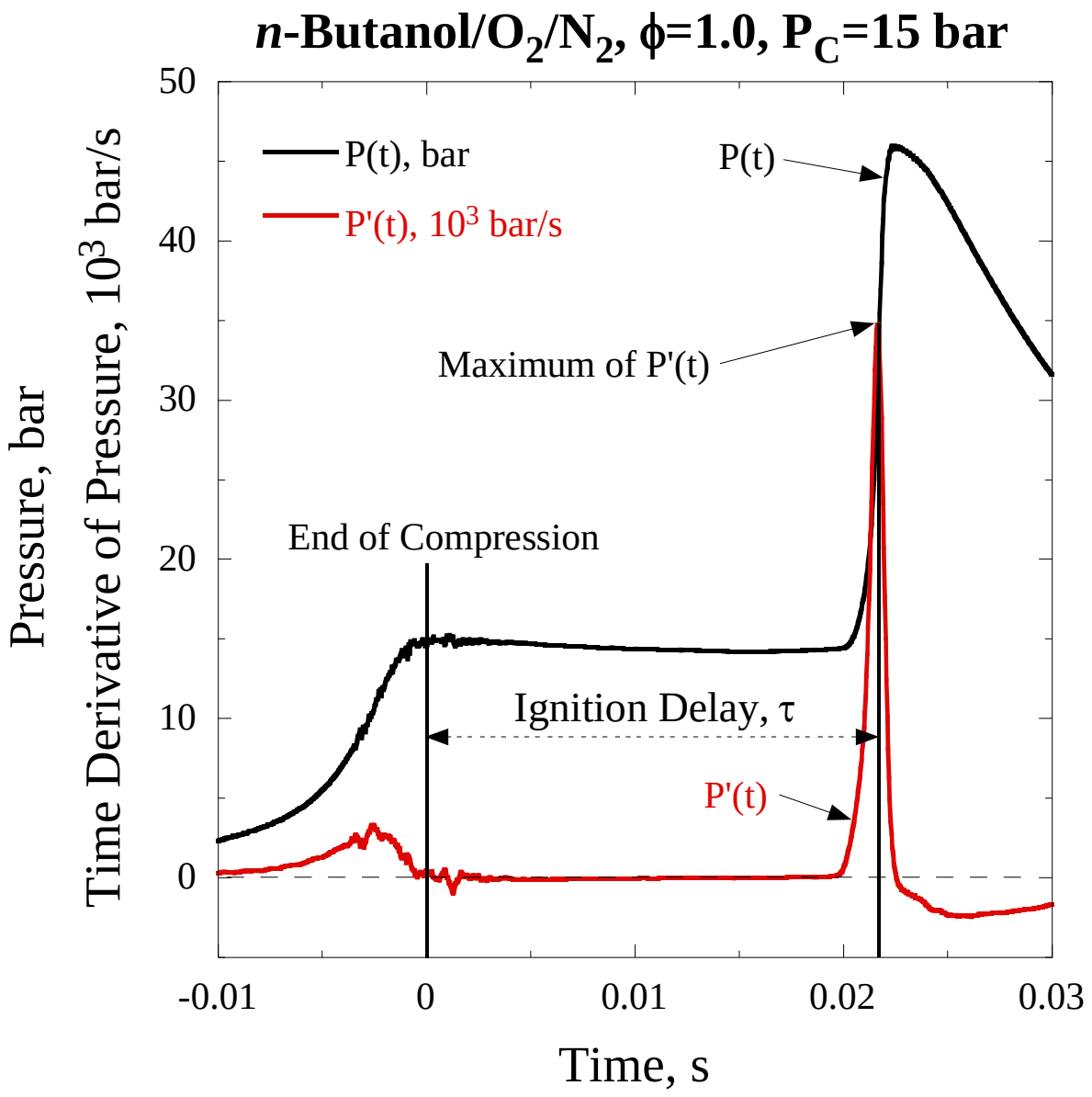

**Figure 2. Definition of ignition delay used in this study. $P'(t)$ is the time derivative of the pressure.**

Temperature at TDC is used as the reference temperature for reporting ignition delay data and is called the compressed temperature ($T_C$). The temperature is calculated using the variable volume simulations. The kinetic mechanisms used in this study are taken from the work by Grana et al.[13], Moss et al.[8], and Van Geem et al.[16]. This approach of deducing $T_C$ requires the assumption of an "adiabatic core" of mixture in the reaction chamber, which is facilitated on the present RCM by the creviced piston discussed previously. To ensure no significant chemical heat release is contributing to the determination of the temperature at TDC, calculations are performed and compared with and without reaction steps for each kinetic mechanism; the temperature profile during the compression stroke is the same whether or not reactions are included. This approach has been validated in Refs. 17 and 21-26.

### G. Definition of Ignition Delay

The end of compression, when the piston reached TDC, is identified by the maximum of the pressure trace ($P(t)$) prior to the ignition point. The local maximum of the derivative of the pressure trace with respect to time ($P'(t)$), in the time after TDC, is defined as the point of ignition. The ignition delay is the time difference between the point of ignition and the end of compression. Figure 2 illustrates the definition of ignition delay ($\tau$) used in this study.

## III. Results and Discussion

Figures 3(a)-3(d) shows the experimental pressure traces for each of the fuels, with each run labeled by its compressed temperature. The non-reactive case, described previously, is a run with oxygen in the mixture replaced by nitrogen to suppress oxidation reactions but maintain a similar specific heat ratio. These plots clearly demonstrate one of the primary advantages of the RCM – namely, the ability to vary compressed temperature while maintaining similar compressed pressure. Each of the fuels has monotonically decreasing ignition delay with increasing temperature, indicating there is no NTC region present in this temperature and pressure range. In addition, there is clearly no evidence of two-stage ignition for any of these fuels under the conditions investigated.

Furthermore, for *tert*- and *iso*-butanol, the pressure traces for the reactive runs closely match the non-reactive case, until the pressure spikes due to hot ignition. On the other hand, for *n*-butanol, prior to hot ignition, there is a slight deviation from the non-reactive case, indicating minor pre-ignition chemical heat release.

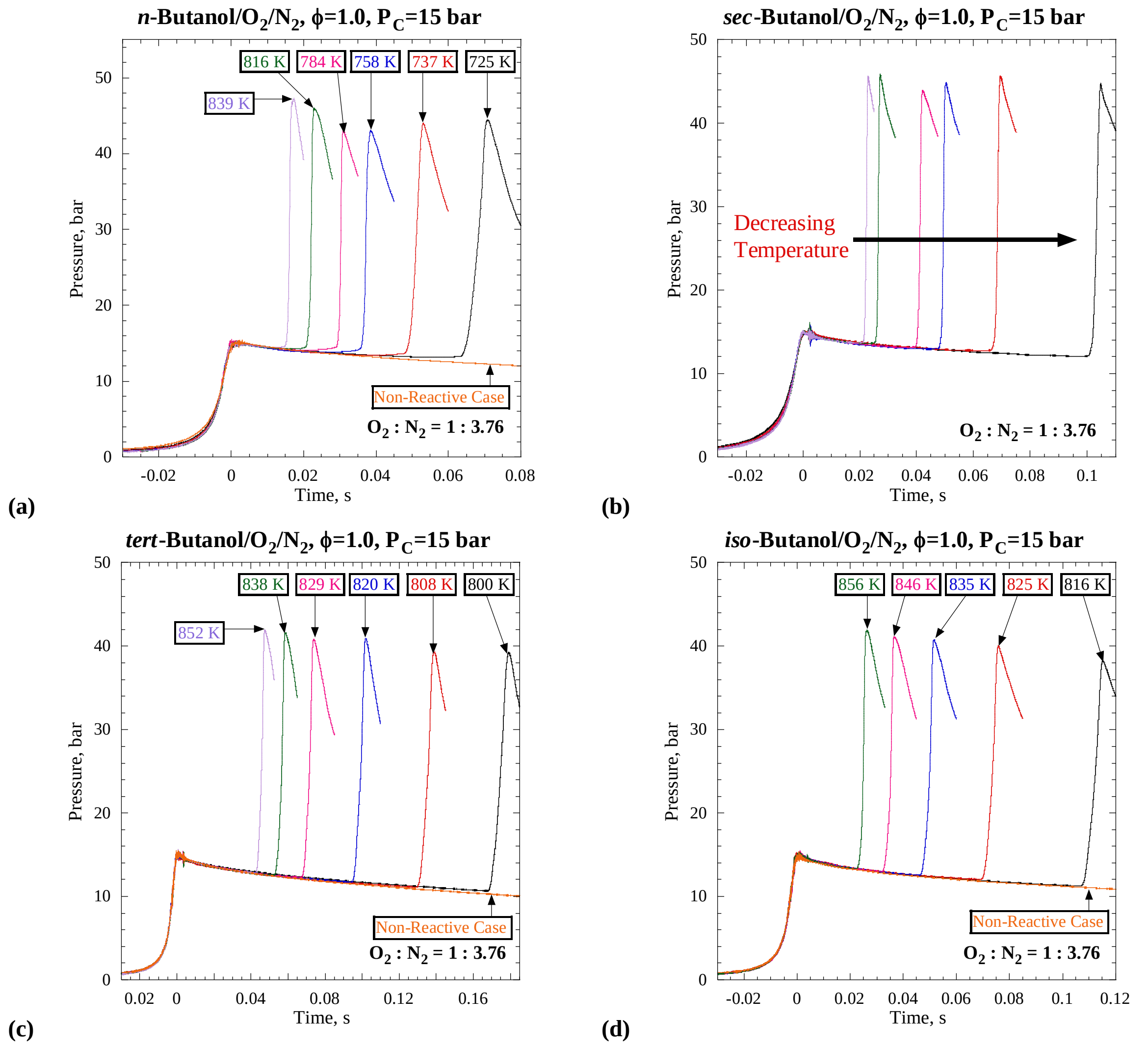

**Figure 3. Experimental pressure traces in the RCM for the four isomers of butanol. Note the absence of NTC and two-stage ignition on these plots.**

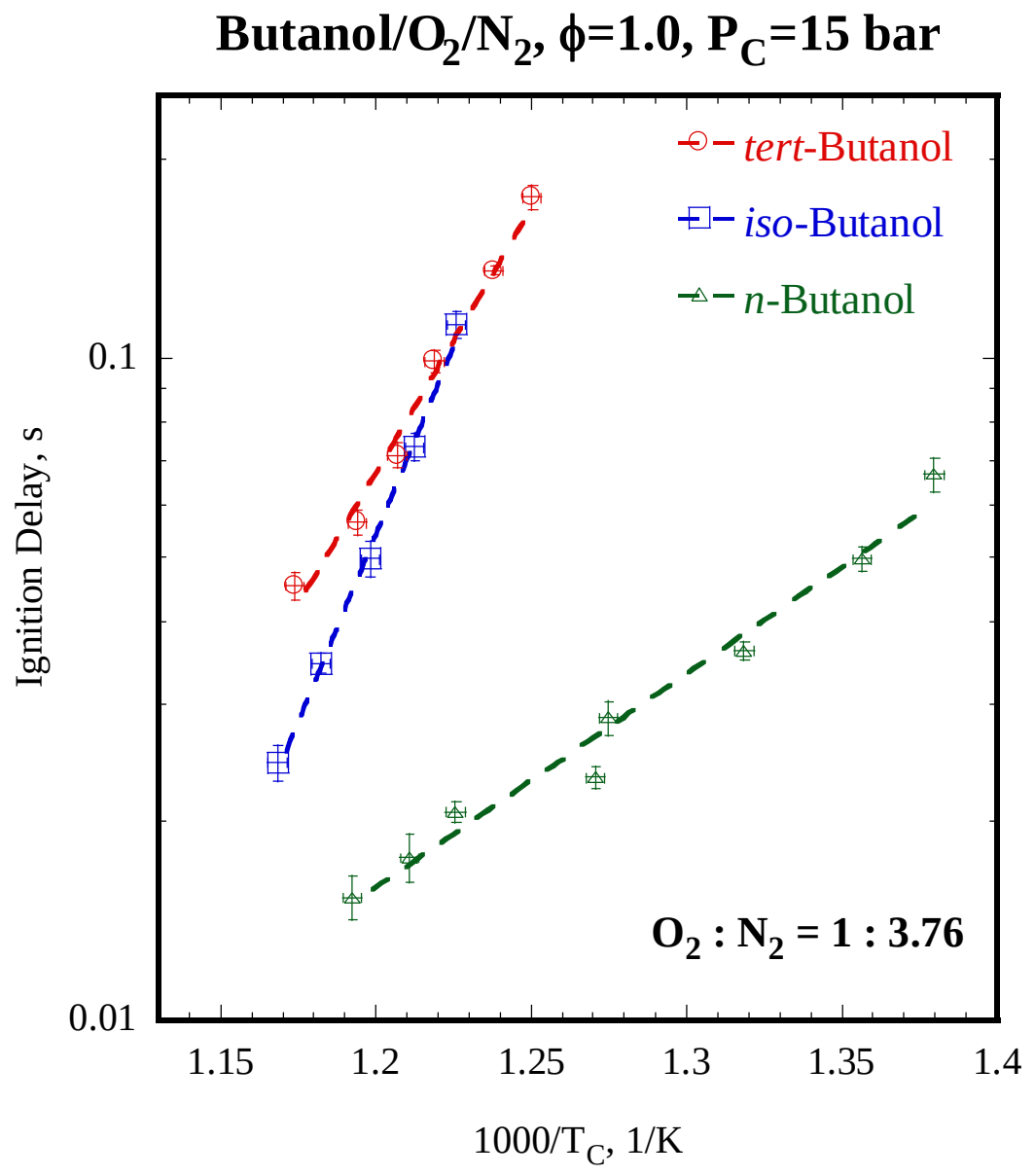

**Figure 4. Arrhenius plot of the ignition delays of *n*-, *iso*-, and *tert*-butanol.**

Figure 4 shows an Arrhenius plot of the ignition delays of *n*-, *tert*- and *iso*-butanol. The vertical error bars represent two standard deviations of the ignition delays, calculated from all the runs at that condition; the horizontal error bars are the uncertainty in the determination of $T_C$ related to thermocouple measurement of the initial temperature and piezoelectric transducer measurement of pressure[23]. The dashed lines are least-squares fits to the data.

There are several interesting features in Fig. 4. As expected, *n*-butanol is the most reactive of the butanol isomers, and *tert*-butanol is the least reactive. Reactivity in terms of ignition delay is generally considered to have an inverse relationship; that is, a shorter ignition delay at a similar temperature implies greater reactivity. This ordering matches the results found in previous studies[8,9,11,12]. This is despite large differences in experimental conditions and even several different types of experiments. For instance, Moss et al.[8] measured high temperature, low pressure ignition of the butanol isomers – for their stoichiometric experiments, their temperature and pressure conditions were 1250–1800 K and 1 bar, respectively. In addition, they were using relatively low fuel concentration (1% by mole), whereas the present study used a higher fuel concentration (3.38% by mole). Still, they found *n*-butanol to be most reactive, *iso*-butanol to have indermediate reactivity, and *tert*-butanol to be least reactive.

The second interesting feature of Fig. 4 is the appearance of a "crossing point", where the ignition delay of *iso*-butanol appears to cross over *tert*-butanol. The crossover appears to occur at approximately 815 K. Unfortunately, the current data set for *iso*-butanol is limited by the physical limits of the current RCM. However, future data sets are planned to extend the data to lower temperatures to investigate this feature more systematically.

Figures 5(a)-5(d) show the ignition delays of the isomers compared to simulations using mechanisms available in the literature. Data points represent the current experiments, with vertical and horizontal error bars indicating two standard deviations and the uncertainty in the compressed temperature, respectively. The dashed lines are least squares fits to the data, the solid lines are constant volume, adiabatic simulations, and when included, the dotted lines are "volume as a function of time" simulations. The mechanisms from Moss et al.[8] and Grana et al.[13] overpredict the ignition delay for *n*-, *tert*- and *iso*-butanol, sometimes by as much as two orders of magnitude. This is not surprising, since neither mechanism includes low-temperature chemistry of the butanols. Therefore, "volume as a function of time" simulations are not computed for either of these mechanisms.

The mechanism from Van Geem et al.[16] overpredicts the ignition delays for *n*-butanol, but underpredicts the ignition delay for *iso*- and *tert*-butanol. However, the simulations are quite close to the experimental values for *n*-, *iso*-, and *tert*-butanol. Including post-compression heat loss in the "volume as a function of time" simulations further improves the predictions for *iso*- and *tert*-butanol. However, "volume as a function of time" simulations for *n*-butanol do not improve the agreement with experiments, and so are not shown. It is also interesting to note that the order of reactivity of the mechanisms differs for some fuels in this temperature and pressure range. For *n*- and *iso*-butanol, the order from most to least reactive is Van Geem et al.[15], Moss et al.[8], Grana et al.[13]. For *sec*-butanol, the order is Moss et al.[8], Van Geem et al.[15], Grana et al.[13]. Finally, for *tert*-butanol, the order is Van Geem et al.[15], Grana et al.[13], Moss et al.[8].

It is useful to make one final point about including heat loss after compression in the "volume as a function of time" simulations. As discussed in conjunction with Fig. 3, there is little to no significant heat release prior to the hot ignition. In cases like this, including heat loss tends to increase the ignition delay. Therefore, we do not include "volume as a function of time" simulations for mechanisms that overpredict the ignition delay anyways.

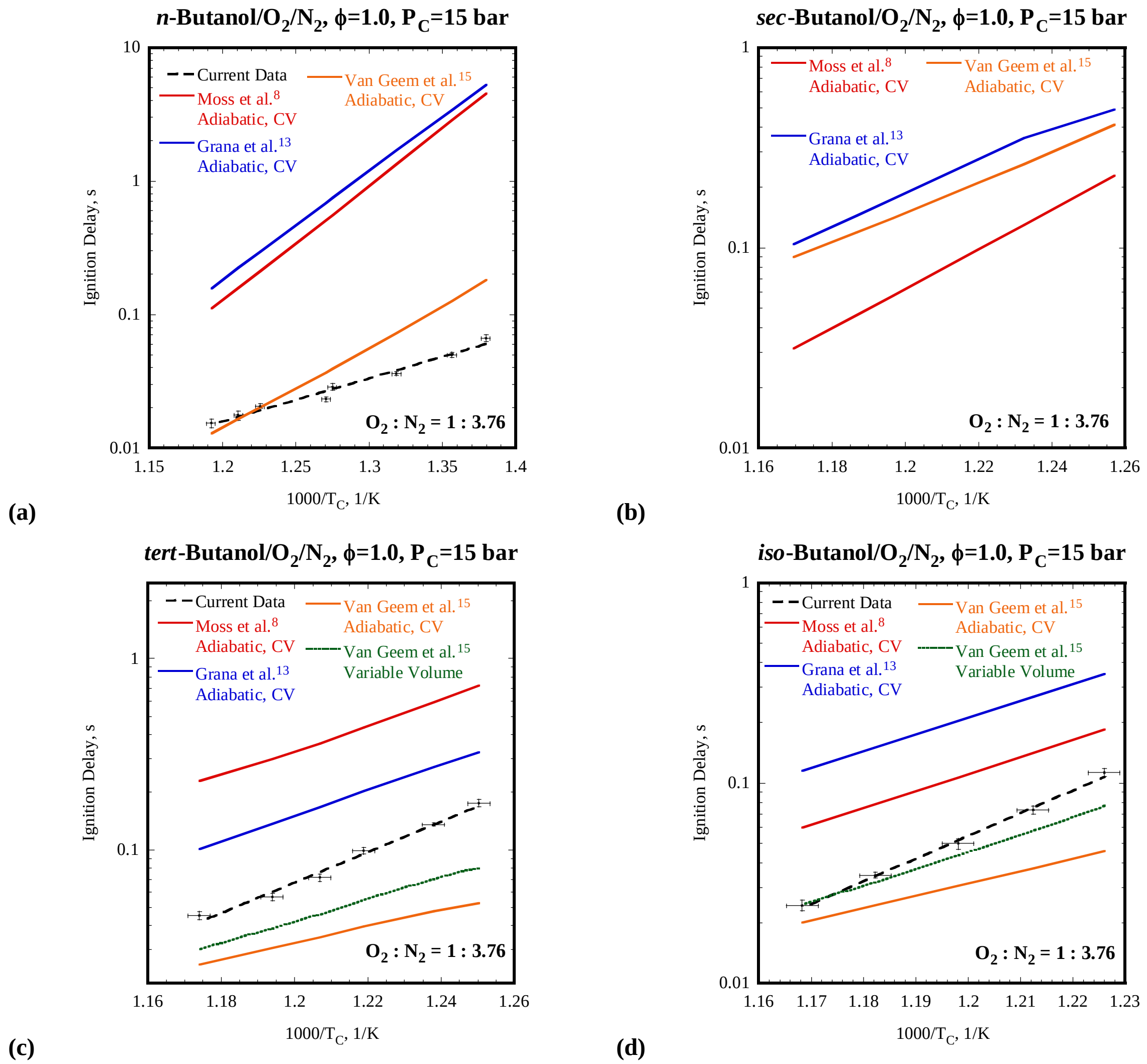

**Figure 5. Arrhenius plots of ignition delays for the four isomers, with simulations.**

## IV. Conclusions

In this rapid compression machine study, autoignition delays of the four isomers of butanol are measured at low temperature and elevated pressure. In particular, compressed temperature conditions from $T_C = 725\text{K} - 855\text{K}$ and compressed pressure condition of $P_C = 15$ bar are studied at equivalence ratio of $\phi = 1.0$ in nitrogen-oxygen air. Of particular note in these experiments is a lack of NTC region and two-stage ignition for all the fuels and conditions studied. The reactivity of the isomers of butanol in this pressure and temperature range is: $n$-butanol > $iso$-butanol > $tert$-butanol, but this ordering appears to be a function of temperature.

Constant volume, adiabatic simulations computed using three kinetic mechanisms available in the literature show a wide variation in agreement with experimental results. In particular, the mechanism of Van Geem et al.[16] shows good agreement with *n*-, *tert*-, and *iso*-butanol. Including the effect of heat loss from the reactants to the reactor walls improved predictions of ignition delay for *tert*- and *iso*-butanol, but not for *n*-butanol.

## Acknowledgments

This material is based upon work supported as part of the Combustion Energy Frontier Research Center and Energy Frontier Research Center funded by the U.S. Department of Energy, Office of Science, Office of Basic

Energy Sciences, under Award Number DE-SC0001198. We thank Dr. Y. Zhang at the University of Connecticut for his help operating the GCMS.